\documentclass[aps, pra, twocolumn, notitlepage, nofootinbib, showkeys]{revtex4-1}

\usepackage{graphicx,amsmath,pstricks,subfig,float}
\begin{document}
	
\title{Long-term implications of observing an expanding cosmological civilization}
	
\author{S. Jay Olson}
\email{stephanolson@boisestate.edu}
\affiliation{Department of Physics, Boise State University, Boise, Idaho 83725, USA}

\date{\today}
	
\keywords{Cosmology, SETI}

\begin{abstract}

Suppose that advanced civilizations, separated by a cosmological distance and time, wish to maximize their access to cosmic resources by rapidly expanding into the universe.  How does the presence of one limit the expansionistic ambitions of another, and what sort of boundary forms between their expanding domains?  We describe a general scenario for any expansion speed, separation distance, and time.  We then specialize to a question of particular interest:  What are the future prospects for a young and ambitious civilization \emph{if they can observe the presence of another} at a cosmological distance?  We treat cases involving the observation of one or two expanding domains.  In the single-observation case, we find that almost any plausible detection will limit one's future cosmic expansion to some extent.  Also, practical technological limits to expansion speed (well below the speed of light) play an interesting role.  If a domain is visible at the time one embarks on cosmic expansion, higher practical limits to expansion speed are beneficial only up to a certain point.  Beyond this point, a higher speed limit means that gains in the ability to expand are more than offset by the first-mover advantage of the observed domain.  In the case of two visible domains, it is possible to be ``trapped" by them if the practical speed limit is high enough and their angular separation in the sky is large enough, i.e. one's expansion in any direction will terminate at a boundary with the two visible civilizations.  Detection at an extreme cosmological distance has surprisingly little mitigating effect on our conclusions.

\end{abstract}

\maketitle

\section{Introduction}
The Hart-Tipler argument is that advanced life must be absent from our galaxy, due to the implications of some plausible technologies, especially self-reproducing spacecraft~\cite{hart1975,tipler1980}.  Such technology would have the power to fully occupy our galaxy on a timescale that is orders of magnitude shorter than the age of the Milky Way, or the timescale for the biological evolution of intelligence~\cite{jones1976,sagan1983,valdes1980}.  A related argument can also be applied at the intergalactic scale, with somewhat different conclusions:  Our existence at the present cosmic time constrains the appearance rate of civilizations that expand between galaxies with self-replicating technology.  We have argued that intergalactic civilizations might exist and be observable, but if so, they would have to be so rare as to almost certainly appear at a cosmological distance from us~\cite{olson2015a}.  

Due to our lack of direct information on the nature of extraterrestrial intelligence, these arguments fall into a particular class of approaches to understanding the Fermi Paradox -- one that assigns high significance to the current lack of local evidence for ETI, and assumes fairly high limits to engineering ability and ambition for life.  Other approaches have long existed, e.g. those discounting the current state of the evidence~\cite{ball1973} or the possibility of expansionistic ambitions~\cite{sagan1983}.  In such alternatives, advanced life may be very prevalent within our own galaxy and others, but very difficult to observe.  This makes the implications of observation and contact very uncertain, which has lead to, for example, debates over the wisdom of METI (Messaging to Extra-Terrestrial Intelligence)~\cite{gertz2016}.  However, if we assume advanced life is rare but (sometimes) expansionistic, the implications of a detection at an intergalactic distance are far more long-term in character, and amenable to geometrical modeling -- a feature that we begin to develop here.

We note that extragalactic SETI is quite new~\cite{annis1999b}, with a surge of recent attempts to find Kardashev type iii civilizations~\cite{kardashev1964} in survey data of nearby galaxies~\cite{wright2014b,griffith2015,zackrisson2015,villarroel2016}, and evidence for even larger-scale technology at a cosmological distance in CMB maps~\cite{lacki2016}.   We have, in turn, argued that a Kardashev type iii civilization would seem to imply all of the technology and ambition required to expand at an intergalactic and eventually a cosmological scale~\cite{olson2015a}.  If this reasoning is valid, then large and expanding domains of life-altered galaxies are actually more likely to be observed than isolated ones\footnote{At present, we take no position on how galaxies are likely to be engineered, or how they would be identified, as there are many possibilities~\cite{lacki2016}.  We are focused here on far simpler questions of large-scale geometry and access to resources, in the context of highly driven civilizations.}.

From an independent field of study, it has been argued that resource acquisition is one of the ``basic drives" of a generic superintelligent AI~\cite{omohundro2008,bostrom2014}.  This means, in essence, that a sufficiently powerful AI will tend to use extreme expansion and resource acquisition as a means of maximizing its utility, unless it is explicitly and carefully designed to avoid such behavior (a \emph{generic} artificial superintelligence is pathological by human standards -- it has a model of itself, a model of the world, and a utility function to maximize through its actions -- it cares about nothing else).  That the behavior of generic superintelligences tend to display common characteristics is known as the ``instrumental convergence thesis" to the research community~\cite{bostrom2012}. This seems to imply that, even if advanced alien species tend to be monks who have forsaken all worldly gain, the \emph{accidents} involving insufficiently careful design of an artificial superintelligence are potentially one of the largest observable phenomena in the universe, when they occur.  The word ``civilization" is not really the best description of such a thing, but we use it for the sake of historical continuity.

Such possibilities raise questions about the future of humanity.  Is an advanced form of humanity, or its successor intelligence, poised to create an expanding domain of engineered (or at least occupied) galaxies~\cite{fogg1988,armstrong2013}?  If so, what are the limits to how much of the universe we could eventually occupy?  One limitation is set by the cosmic event horizon -- even traveling arbitrarily close to speed of light, only galaxies within a finite co-moving distance are reachable.  The limiting factor we explore here is the presence of other expanding civilizations, consisting of galaxies already fully saturated with alien technology.  In the cosmological model of aggressively expanding life~\cite{olson2014}, one can calculate average final domain volumes that account for these limitations, as a function of the starting time and the appearance rate of expanding civilizations.  But universe-averaged quantities do not tell us much about the geometry of a specific situation, or what is implied by observing an expanding civilization.

Here, we study and illustrate the geometry of colliding domains of expanding civilizations, belonging to distinct species who do not wish to share (or fight over) resources.  We use the results to describe what is implied for the possible future of humanity, if a rapidly expanding cosmological civilization is discovered in present-day observations.  The underlying assumptions, enumerated in the next section, are as simple as possible, forming a framework on which more elaborate behavior modeling and game-theoretic considerations might be incorporated in future work.  In section II, we describe the boundary that forms between two expanding civilizations as a function of expansion speed, separation distance, and starting times.  The boundary takes the form of a hyperboloid in co-moving coordinates, and we give expressions for the final co-moving volume of each civilization.  In section III we specialize to our main question of interest -- what are the long-term implications of actually \emph{observing} a rapidly expanding civilization in extragalactic SETI?  We illustrate how results depend on the practical limit to expansion velocity, $v$, and  more weakly on the separation distance, and that there exists an optimum value of $v$ for the observing civilization.  Above the optimum value, gains in the ability to rapidly expand are more than offset by the increasingly dominant presence of the observed civilization.  In section IV, we study implications of observing two expanding civilizations, depending on their angular separation in the sky.  In scenarios where a high expansion speed is practical, it becomes likely that the observer is ``trapped" if two expanding civilizations are observable (i.e. one will eventually reach a boundary with the two visible civilizations when expanding in any direction).  Section V contains a discussion of these results, and our concluding remarks.  We include an appendix that illustrates how closely our homogeneous expansion model approximates a branching expansion model in the context of discrete galaxies, realistic cosmic structure, and plausible technologies.

\section{Geometry of two colliding domains}

Our results will be derived in the context of a completely homogeneous cosmology, with domains expanding according to a simple geometrical rule, which we now describe in the following list of assumptions and notation.  This is an approximation by necessity, but as we illustrate in the appendix (by considering a map with over 40,000 galaxies), it accurately reflects a simple and plausible expansionistic behavior in the context of realistic cosmic structure.

\begin{enumerate}
	\item Spacetime is assumed to be a spatially flat and homogeneous Friedmann-Robertson-Walker (FRW) solution\footnote{For numerical calculations, we assume a solution with $ \Omega_{\Lambda 0} = .683$, $\Omega_{r 0} = 3 \times 10^{-5} $, $\Omega_{m 0} = 1-\Omega_{\Lambda 0} - \Omega_{r 0} $, and $H_0 = .069 \ Gyr^{-1} $, so that the present age of the universe is $t_{0} \approx 13.75$ Gyr.}.  The scale factor is denoted by $a(t)$, with  $a(t_0) = 1$.  We use co-moving coordinates, so that the metric reads $ds^2 = -dt^2 + a(t)^2 (dx^2 + dy^2 + dz^2)$, and we use units such that $c=1$.  All references to distances and volumes refer to co-moving distance and volume.
	\item In this homogeneous framework, the first civilization to reach a point in space is assumed to own its resources entirely, leading to a thin boundary between civilizations.  This is motivated by considering spacecraft that reproduce exponentially upon reaching a new galaxy, quickly establishing a dominating presence throughout it.  Because theoretical colonization times with self-reproducing spacecraft are shorter than intergalactic travel times~\cite{jones1976}, we do not expect an extended ``thick" boundary, and only a rare shared galaxy at the boundary.
	\item The expanding civilizations are taken to originate at coordinates $ \{ x_1, y_1, z_1 \} = \{-C, 0, 0\}$ and $\{x_2, y_2, z_2\} = \{C, 0, 0\}$, i.e. the co-moving distance between them is taken to be $2C$. $2C$ is assumed to be a cosmological distance, i.e. larger than the homogeneity scale of the universe ($\approx .25$ Gly comoving).
	\item The first civilization begins expanding at cosmic time $t = t_1$, while the second civilization begins expanding at the present cosmic time, $t = t_0$.  Be aware, this means $t_1 < t_0$.
	\item The civilizations expand outward in all directions with a constant velocity $v$ in the co-moving frame of reference (i.e. stationary observers attached to galaxies would see the wave of expansion pass by at speed $v$ in their own local frame of reference).   This model is consistent with a number of aggressive expansion strategies (see the appendix as well as~\cite{olson2014}), and it reflects the assumption that both civilizations will achieve the same practical limits to expansion speed.  The co-moving distance from each origination point to its respective frontier at time $t$ is given by $r_1(t) = \int_{t_1}^{t} \frac{v}{a(t')} dt'$ and $r_2(t) = \int_{t_0}^{t} \frac{v}{a(t')} dt'$.\footnote{The expansion of a spherical light pulse is a special case of this model, with $v = 1$. This is very nearly as simple as constant-$v$ expansion in flat spacetime, where in that case $r(t) = \int_{t_{start}}^{t} v \, dt'$, but the factor of $\frac{1}{a(t')}$ in the integrand implies finite maximum expansion distance of $r(\infty)$ in the standard cosmology.  When $v=1$, the distance $r(\infty)$ corresponds to the cosmic event horizon.} It is assumed that $r_1(t_0) < 2 C$, i.e. that the first civilization has not overtaken the second one at the time it begins to expand.
\end{enumerate}

\begin{figure}[t]
	\centering
		\includegraphics[width=0.8\linewidth]{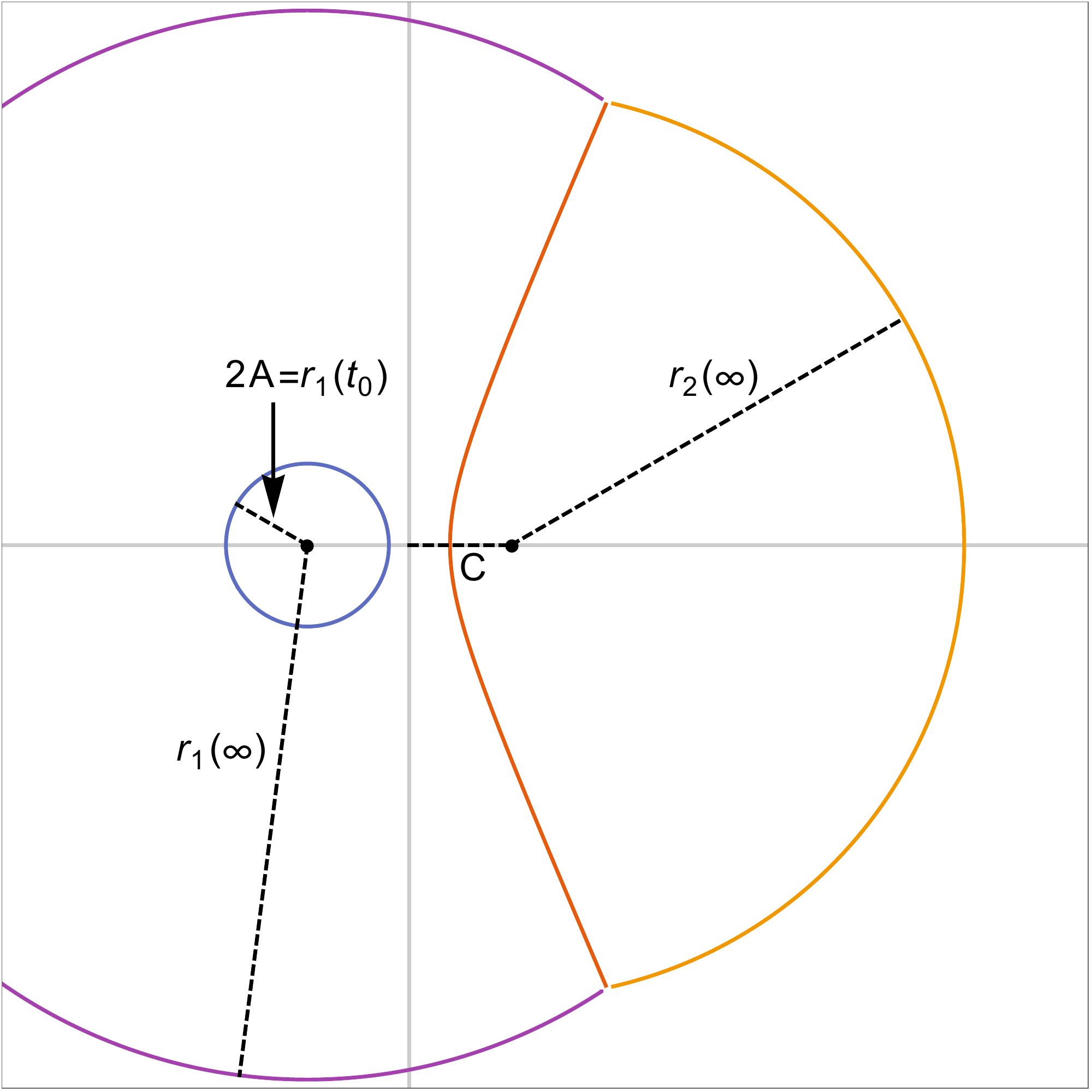}
	\caption{Dimensions of a cosmic-scale collision of expanding civilizations:  At time $t_0$, civ 1 has expanded to radius $2 A$, when civ 2 begins its own expansion a comoving distance of $2 C$ away.  At $t=\infty$, the civilizations will have expanded to comoving radii of $r_1(\infty)$ and $r_2(\infty)$, with a hyperboloid forming the boundary between them. }
\end{figure}

At time $t_0$, an expanding sphere of radius $r_1(t_0)$ has formed about the first civilization, while the second civilization is just beginning their own expansion.  As the spheres expand, they meet at the domain boundary, which is equally distant from point $\{C, 0, 0 \}$ and the sphere of radius $r_1(t_0)$ about point $\{-C,0,0\}$.  This can be expressed as the points satisfying $r_1 - r_2 = r_1(t_0)$, which defines a hyperboloid with foci at the origination points $\{-C, 0, 0 \}$ and $\{C, 0, 0 \}$, and with semi-major axis $A = \frac{1}{2} r_1(t_0)$ -- the domain boundary is the $x>0$ sheet of this hyperboloid (see Figure 1).  

Defining $B^2 \equiv C^2 - A^2$, we express the hyperboloid in canonical form:
\begin{eqnarray}
& &\frac{x^2}{A^2} - \frac{y^2}{B^2} - \frac{z^2}{B^2}  =  1.
\end{eqnarray}

When bounded by the maximum expansion distance at $r_2(\infty)$ from $\{ C,0,0 \}$, the region inside the $x>0$ sheet of the hyperboloid represents the final co-moving volume of space occupied by the second civilization.  This region may be integrated to give the final volume $V_2$ occupied by the second civilization:
\begin{eqnarray}
\resizebox{.42 \textwidth}{!}{$
V_2 = \frac{\pi  (C-A) \left(3 \, r_2(\infty)^2 (C+A)-(C-A)^2 (C+A)+2 \, r_2(\infty)^3\right)}{3 C}. $}
\end{eqnarray}
The volume of the first and larger civilization, $V_1$, is the volume of a sphere, minus that cut out by the hyperboloid:
\begin{eqnarray}
\resizebox{.42 \textwidth}{!}{$
V_1 = \frac{\pi  (C+A) \left(3 \, r_1(\infty)^2 (C-A) - (C-A) (C+A)^2+2 \, r_1(\infty)^3\right)}{3 C}. $}
\end{eqnarray}
These equations are valid for $2C \leq r_1(\infty) + r_2(\infty)$.  At greater separation distance, the civilizations never meet and the final volume is just that of a sphere. 

\section{Implications of observing an expanding alien domain}
We now specialize to the main case of interest.  Suppose civ 2 is a young and ambitious technological species, perhaps comparable to humanity, who is close to embarking on rapid expansion into the universe.  A short time prior to their launch of self-reproducing spacecraft, they perform a detailed galaxy survey and observe civ 1 at a very early stage of expansion, some cosmological distance away.  What does this mean for civ 2's future ambitions?  

This scenario imposes the relation $\int_{t1}^{t_0} \frac{1}{a(t')} \, dt' = 2 C $ -- i.e. between their starting times, light has had just enough time to travel from civ 1 to civ 2.  In this intervening time, civ 1 has also become much larger -- the radius of civ 1 at time $t_0$ is $\int_{t1}^{t_0} \frac{v}{a(t')} \, dt' = 2 A $, though civ 2 is unable to see this directly in their galaxy survey.  Together, these relations mean that $A = v C$, meaning that we only need to specify the separation distance $2 C$ and the expansion velocity $v$, and we can immediately use the results from the previous section to find final volumes.  

Figure 2 illustrates the final geometry of this situation for a separation distance $2C = 3 \, \text{Gly}$ with expansion velocities of $v=.3$, $v=.6$, $v=.9$, and $v=.99$.  We can substitute $A = v C$ into the equations of the previous section to get expressions for the final volumes:
\begin{eqnarray}
\resizebox{.42 \textwidth}{!}{$
V_1 = \frac{1}{3} \pi  (1+v) \left(3 C v^2 X_1^2 (1-v)+2 v^3 X_1^3-C^3 (1-v) (1+v)^2\right)  $}\\ 
\resizebox{.42 \textwidth}{!}{$
V_2 = \frac{1}{3} \pi  (1-v) \left(3 C v^2 X_2^2 (1+v)+2 v^3 X_2^3-C^3 (1-v)^2 (1+v)\right) .  $}
\end{eqnarray}

In order to make the $v$-dependence explicit in the above equations, we have introduced $X_1 = \int_{t_1}^{\infty} \frac{1}{a(t')} \, dt'$ and $X_2 = \int_{t_0}^{\infty} \frac{1}{a(t')} \, dt'$, so that $r_1(\infty) = v \, X_1$ and $r_2(\infty) = v \, X_2$.  Note that $X_1 = X_2 + 2 C$ in this scenario, so that the only place numerical calculations enter is in calculating $X_2$, which is just the event horizon distance at the present time -- for the cosmological parameters we have used, $X_2 \approx 16.7$ Gly.  As before, these equations are valid so long as $2 C \leq r_1(\infty) + r_2(\infty)$, so that the civilizations actually meet.

\begin{figure}[t]
	\centering
	\subfloat[]{
		\includegraphics[width=0.47\linewidth]{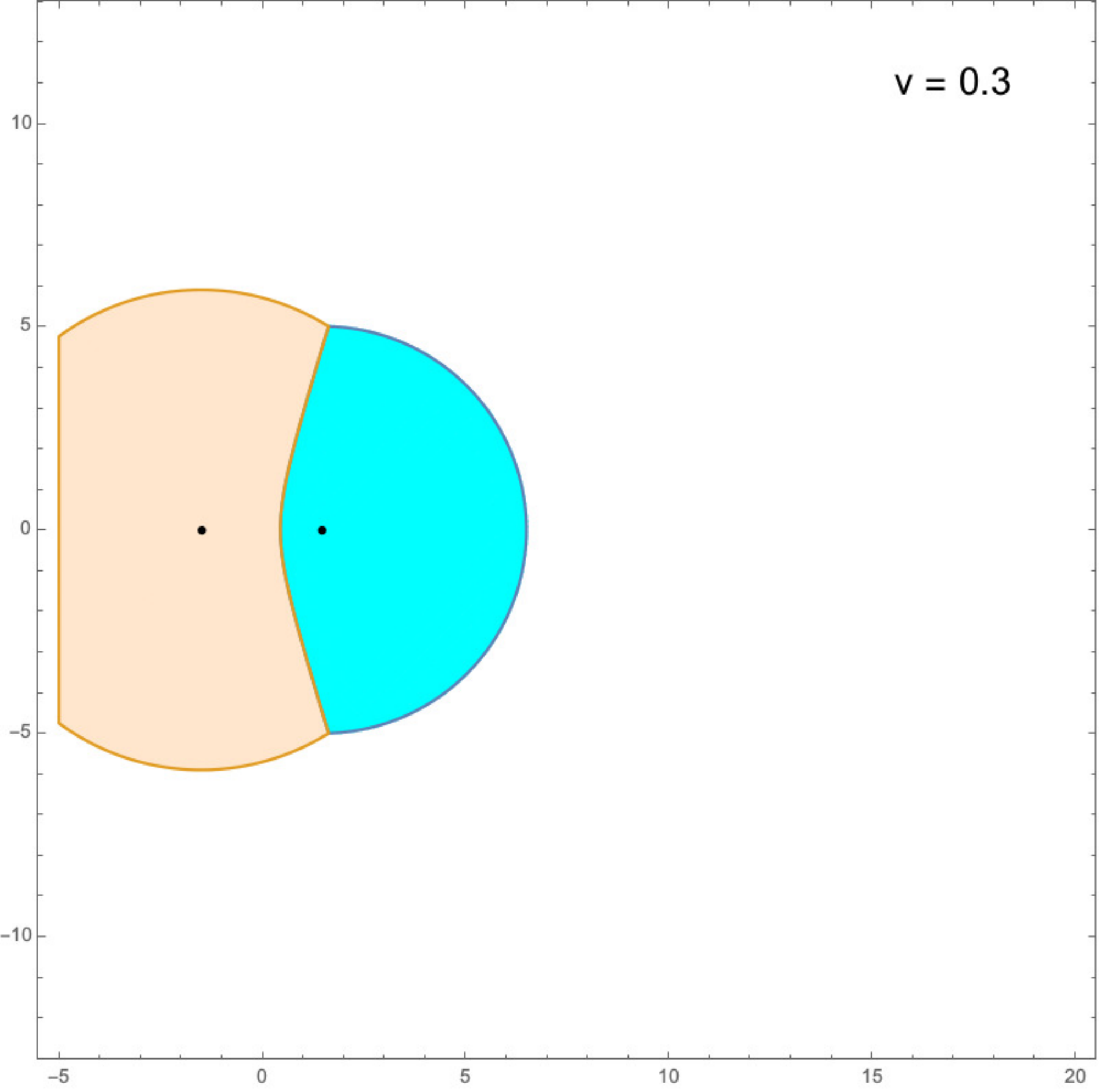}
	}
	\subfloat[]{
		\includegraphics[width=0.47\linewidth]{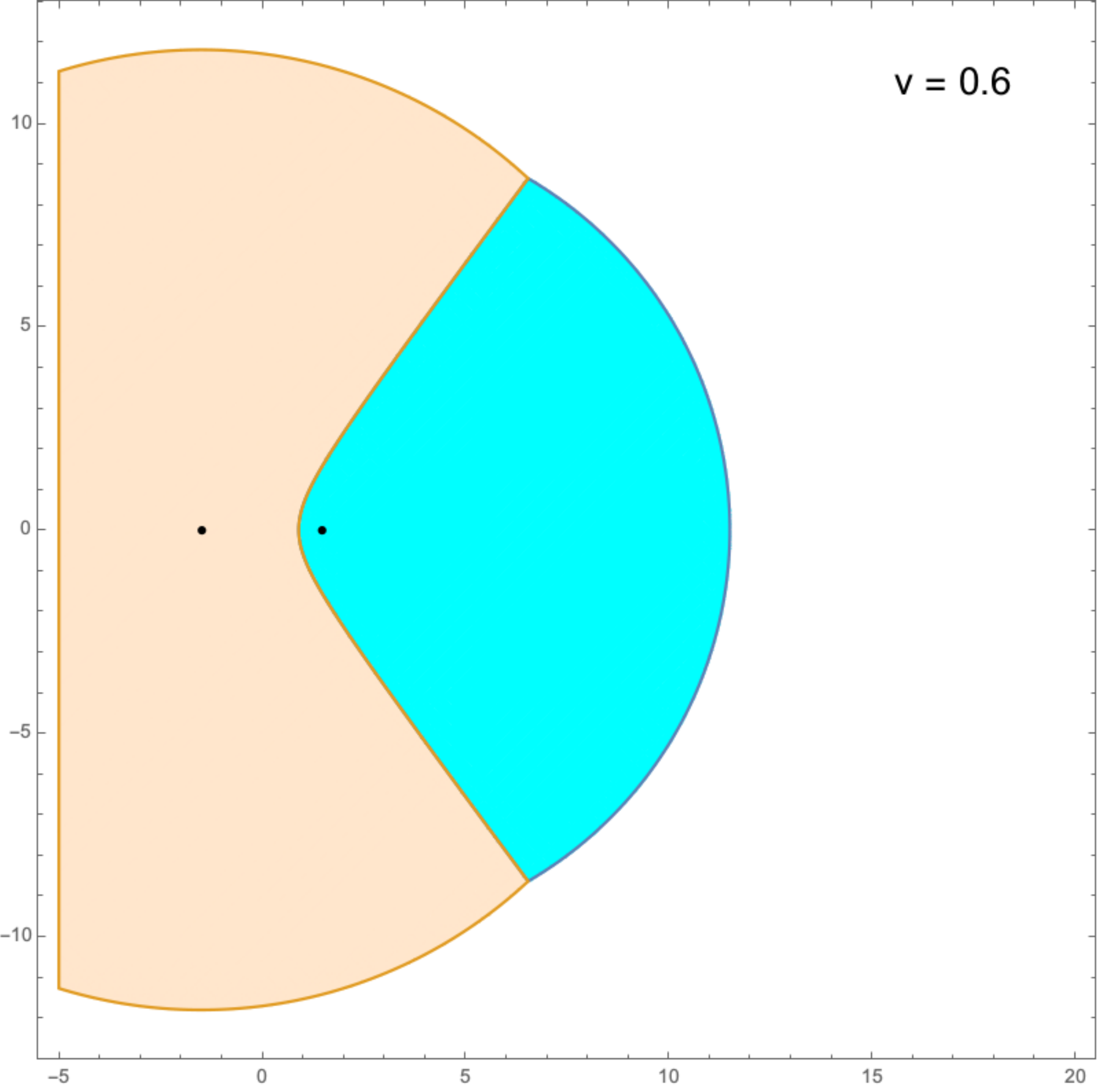}
	}
	\hspace{0mm}
	\subfloat[]{
		\includegraphics[width=0.47\linewidth]{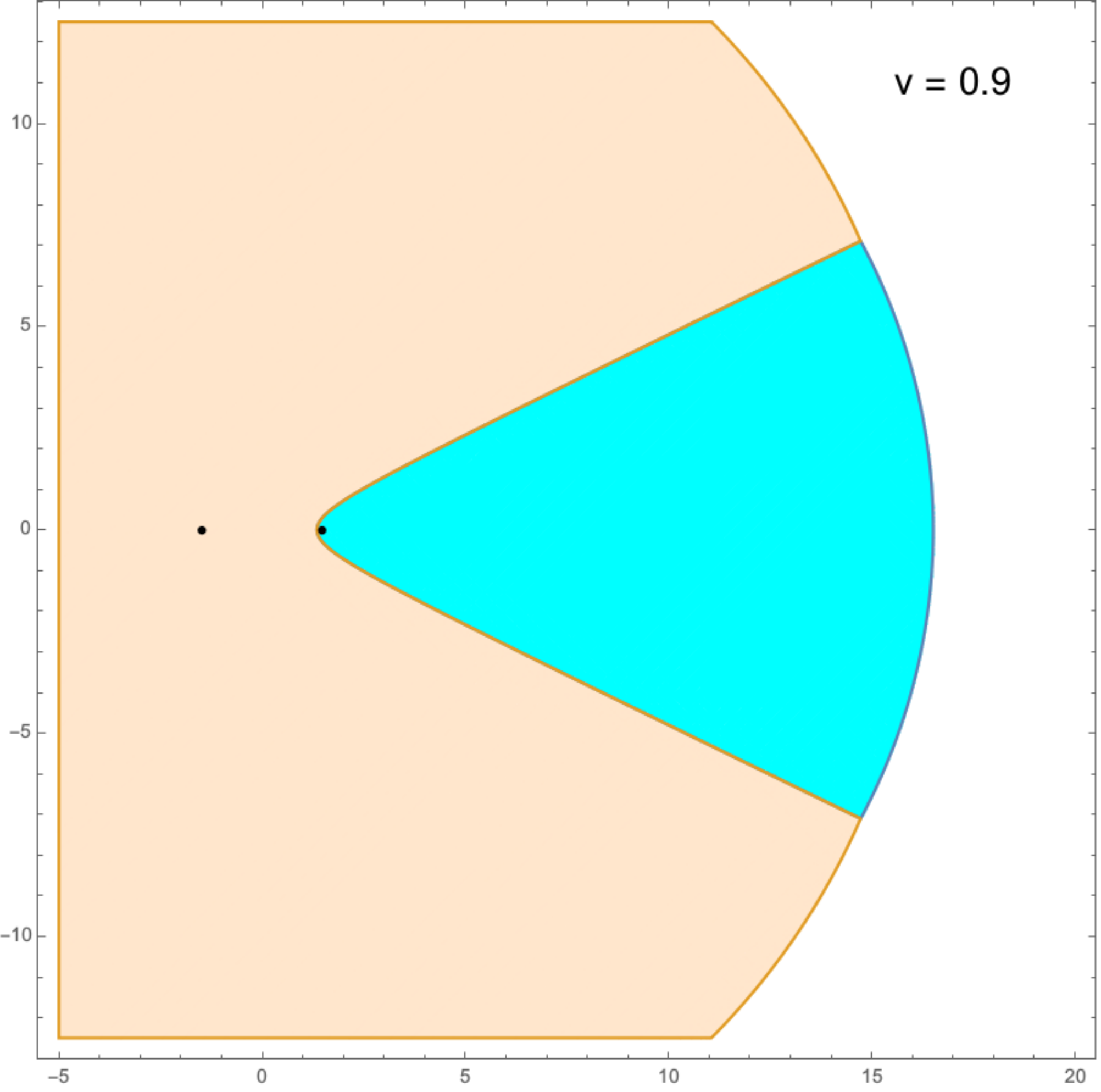}
	}
	\subfloat[]{
		\includegraphics[width=0.47\linewidth]{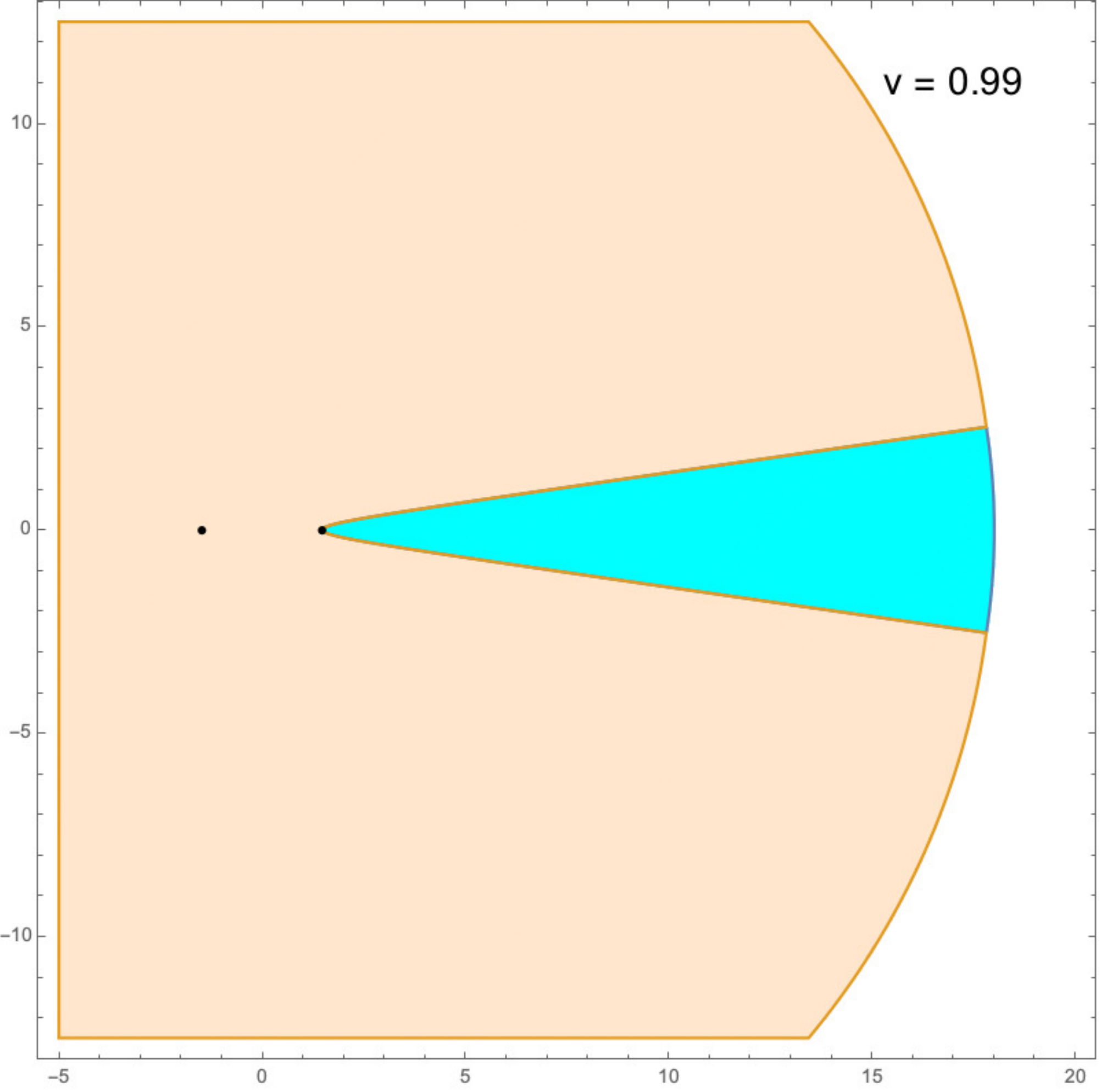}
	}
\caption{Final geometry of four scenarios in which civ 2 (blue) observes civ 1 (yellow) at the earliest stages of expansion and simultaneously begins its own expansion.  Separation distance is taken to be 3 Gly, and the  civilizations have a common expansion speed of: .3 in (a), .6 in (b), .9 in (c), and .99 in (d).  Axes are in units of Gly.}
\end{figure}

Figure 2 shows that the consequences for civ 2 will depend heavily on the maximum practical speed of expansion.  We make this more explicit in Figure 3a, which plots $V_2$ as a function of expansion speed $v$.  In absolute terms, note that scenarios with higher practical limits to technology (higher $v$) give a greater final volume, up to some maximum, whereupon higher limits to technology actually reduce the second civ's final volume, due to the increasingly dominant presence of the first civ.

We can also examine $V_2$ relative to the final volume civ 2 \emph{would have occupied} in the absence of civ 1 -- we call this quantity $V_2 (\mathrm{Competitor} ) / V_2 (\mathrm{No \: Competitor})$.  This fraction is plotted in Figure 3b.  Notice that for sufficiently small $v$, the fraction is exactly 1 -- this corresponds to an expansion speed slow enough that the two civs are never able to meet, though they eventually see one other.

\begin{figure}[t]
	\centering
	\subfloat[]{
		\includegraphics[width=0.9\linewidth]{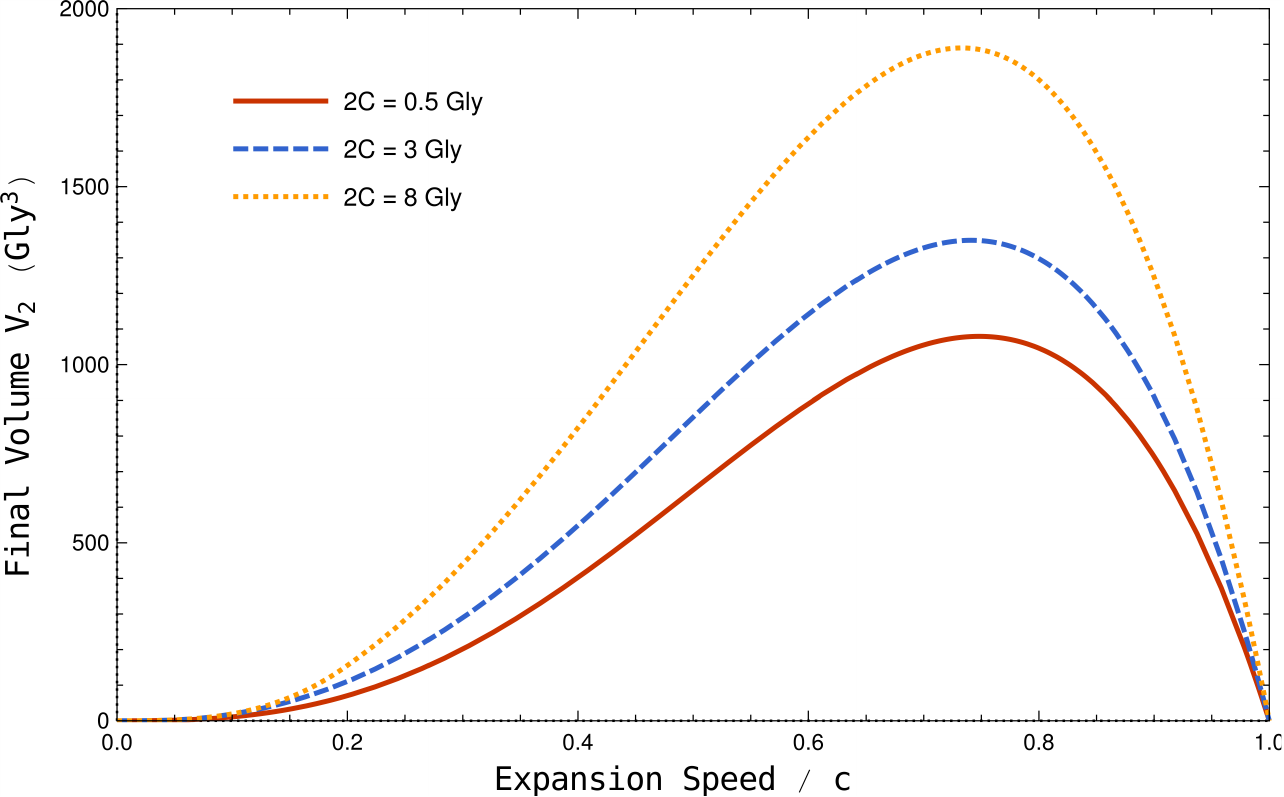}
	}
	\hspace{0mm}
	\subfloat[]{
		\includegraphics[width=0.9\linewidth]{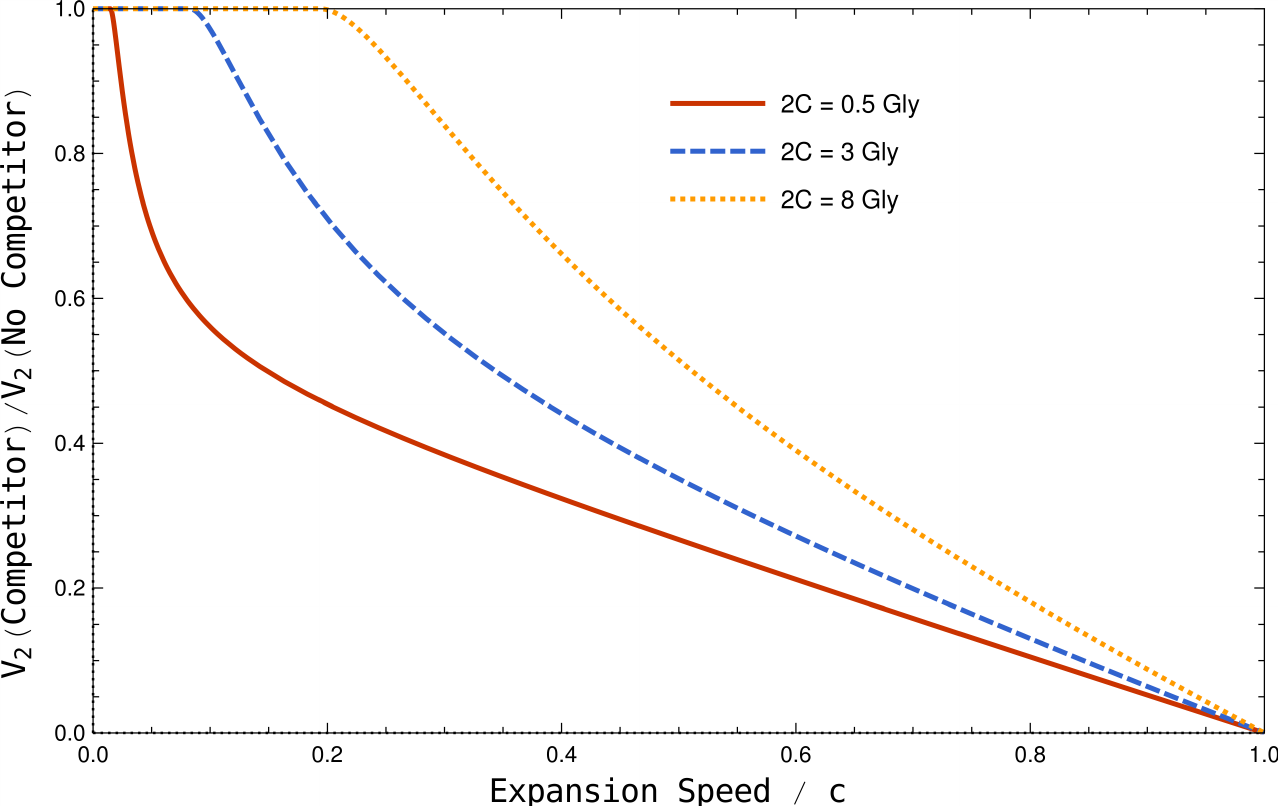}
	}
	\caption{Dependence of final volume $V_2$ on expansion speed.  (a) Higher practical limits to technology (higher $v$) are only beneficial to the observing civilization up to a certain point.  (b) As a fraction of the volume the observing civ could occupy without competition, higher practical speed limits always correspond to greater diminishment by the observed civilization.  }
\end{figure}

Also of interest is the dependence on the separation distance, $2 C$.  As shown in Figure 4, dependence on the separation distance is surprisingly weak over realistic distances\footnote{Based on models for Earthlike planet formation rates and the timescale for biological evolution, observing a civilization beyond co-moving $\approx 8 \, Gly$ seems unlikely, corresponding to a time before the conditions for advanced life to appear were met~\cite{olson2015a,olson2016a}.}, particularly for high-$v$ scenarios.  The difference between observing an expanding civilization just outside our own supercluster and observing one at the maximum plausible distance (8 Gly) amounts to less than a factor of 2 in final volume over a wide range of expansion speeds.

\begin{figure}[h]
	\centering
		\includegraphics[width=0.9\linewidth]{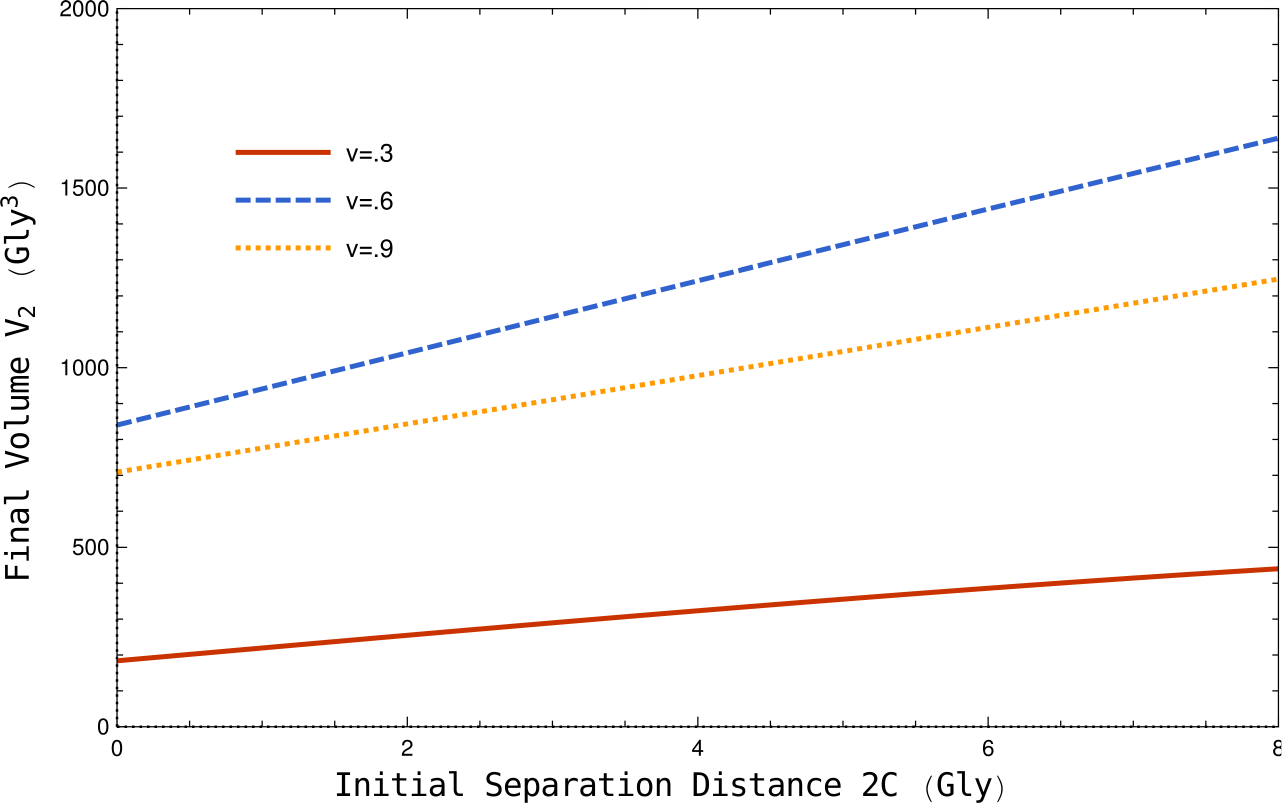}
	\caption{Dependence of final volume $V_2$ on the initial separation distance, $2 C$.  From ``next door" to $8$ Gly, the distance to the observed civilizations amounts to only a factor of $\approx 2$ in final volume to the observing civilization. }
\end{figure}

\section{Two visible expanding domains}

In the case that two expanding domains are observed, the observing civ is left with a region that is the intersection of the interior of two hyperboloids, and is less convenient for obtaining simple volume equations (we use numerical integration to find the final volume).  However, a relevant phenomenon -- trapping -- is easy to analyze.  We describe a civilization as ``trapped" if they reach a domain boundary with another civilization in every direction of their expansion, i.e. they cannot reach their maximum expansion radius of $r(\infty)$ in any direction.  It is very possible that observing two early-stage expanding domains at a cosmological distance implies being trapped by them.  Figure 5a and 5b illustrate two non-trapped scenarios, while Figures 5c represents a critically trapped scenario and 5d is a trapped scenario -- the only difference in these cases is the expansion velocity.

\begin{figure}[t]
	\centering
	\subfloat[]{
		\includegraphics[width=0.47\linewidth]{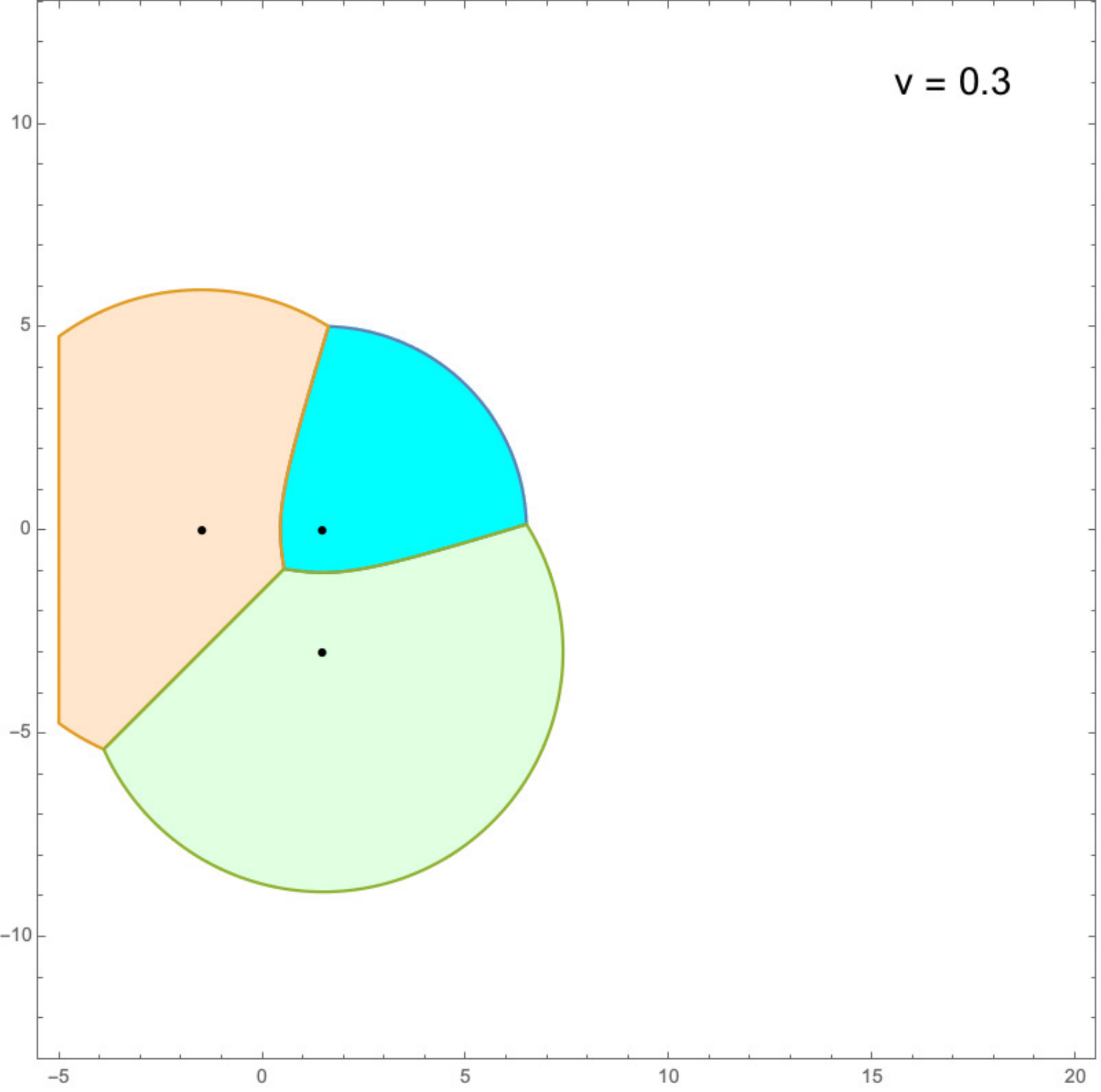}
	}
	\subfloat[]{
		\includegraphics[width=0.47\linewidth]{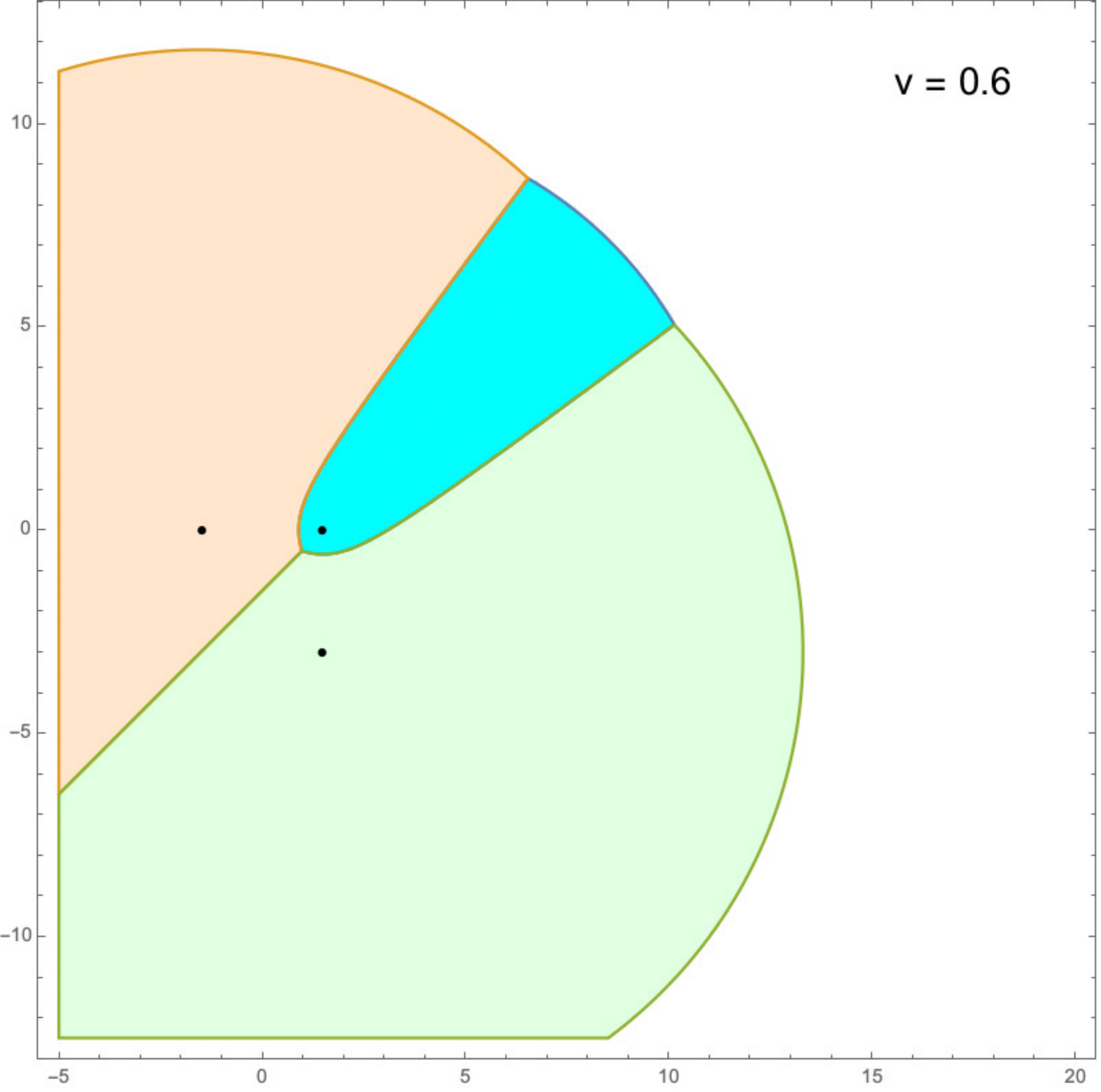}
	}
	\hspace{0mm}
	\subfloat[]{
		\includegraphics[width=0.47\linewidth]{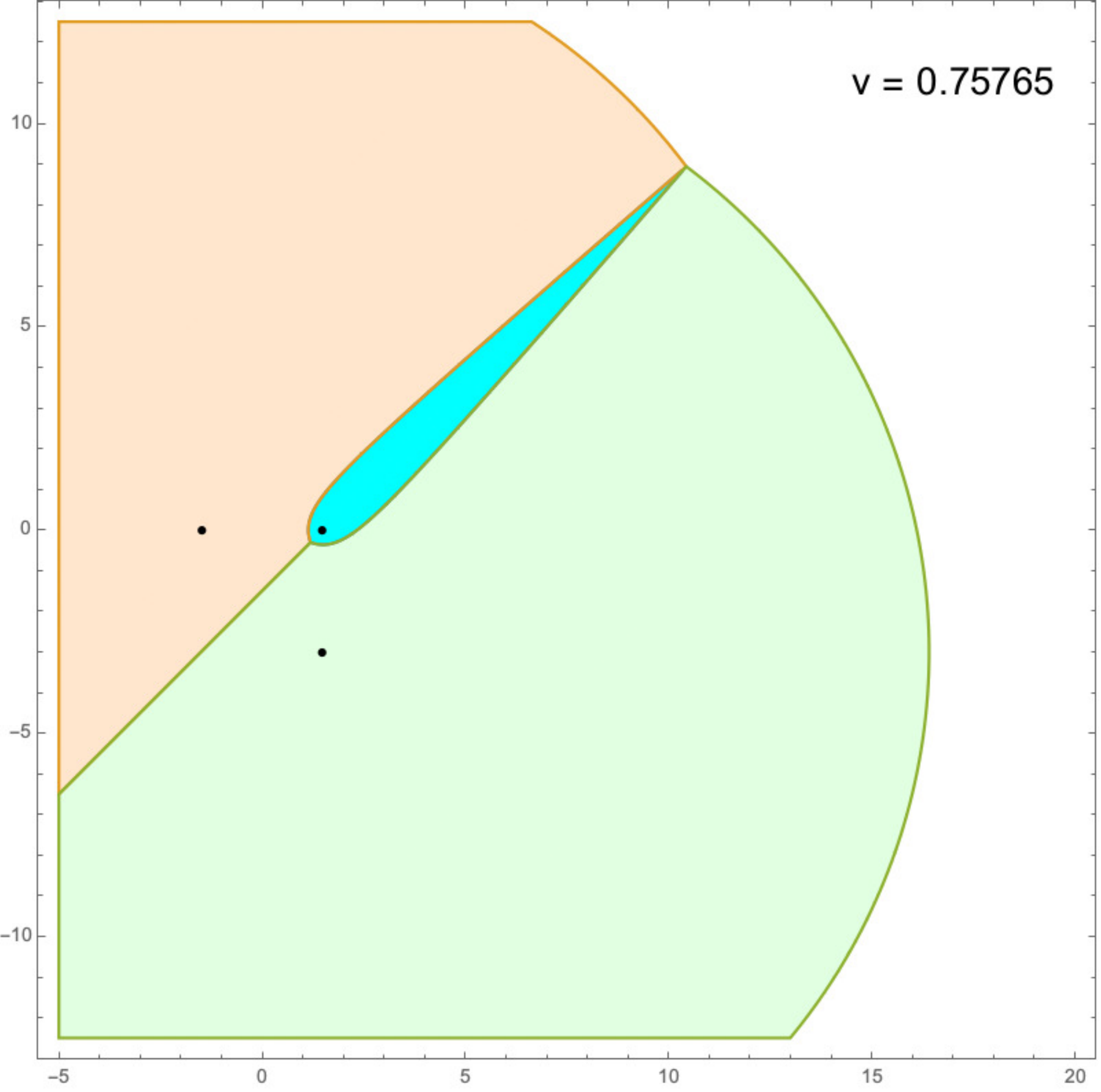}
	}
	\subfloat[]{
		\includegraphics[width=0.47\linewidth]{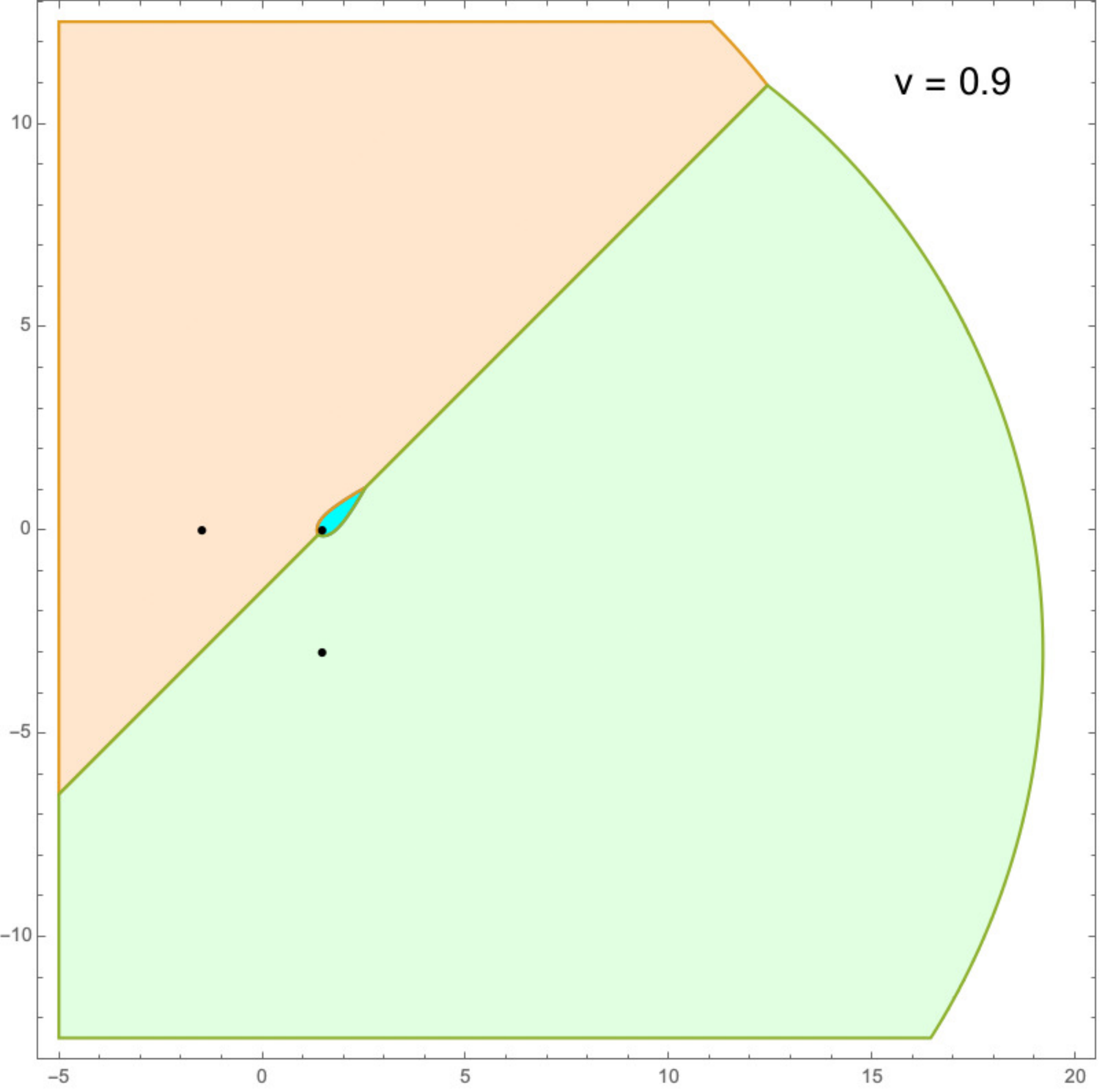}
	}
	\caption{Final geometry of four scenarios in which civ 2 (blue) observes civs 1a and 1b just as it begins expansion.  Angular separation of civs 1a and 1b is $90^{\circ}$, and the initial distance to 1a and 1b is 3 Gly.  Scenarios (a) and (b) are non-trapped, while (c) is critically trapped (at $v = .75765$), and (d) (at $v = .9$) is trapped.  Civ 2 reaches its greatest expansion distance in the plane of the three civilizations, which is shown here.}
\end{figure}

We will assume that two early civilizations (civs 1a and 1b) are visible to civilization 2 at the earliest stages of their expansion and that they appear at the same distance from civ 2.  We will work in the plane formed by the civs, as civ 2 will reach its maximum expansion distance in this plane.  For a given separation of $2 C$ (from civ 2) and angular separation $\theta$ between 1a and 1b (as viewed by civ 2), we want to know the expansion speed above which civ 2 will be trapped.  The critical speed $v_{trap}$ occurs when the three spheres of radii $r_{1a}(\infty)$, $r_{1b}(\infty)$, and $r_{2}(\infty)$ all intersect one another at a single point -- i.e. civ 2 has just barely managed to reach $r_{2}(\infty)$ when further progress would be cut off anyway by civs 1a and 1b (see Figure 5c).  

\begin{figure}[t]
	\centering
	\includegraphics[width=0.9\linewidth]{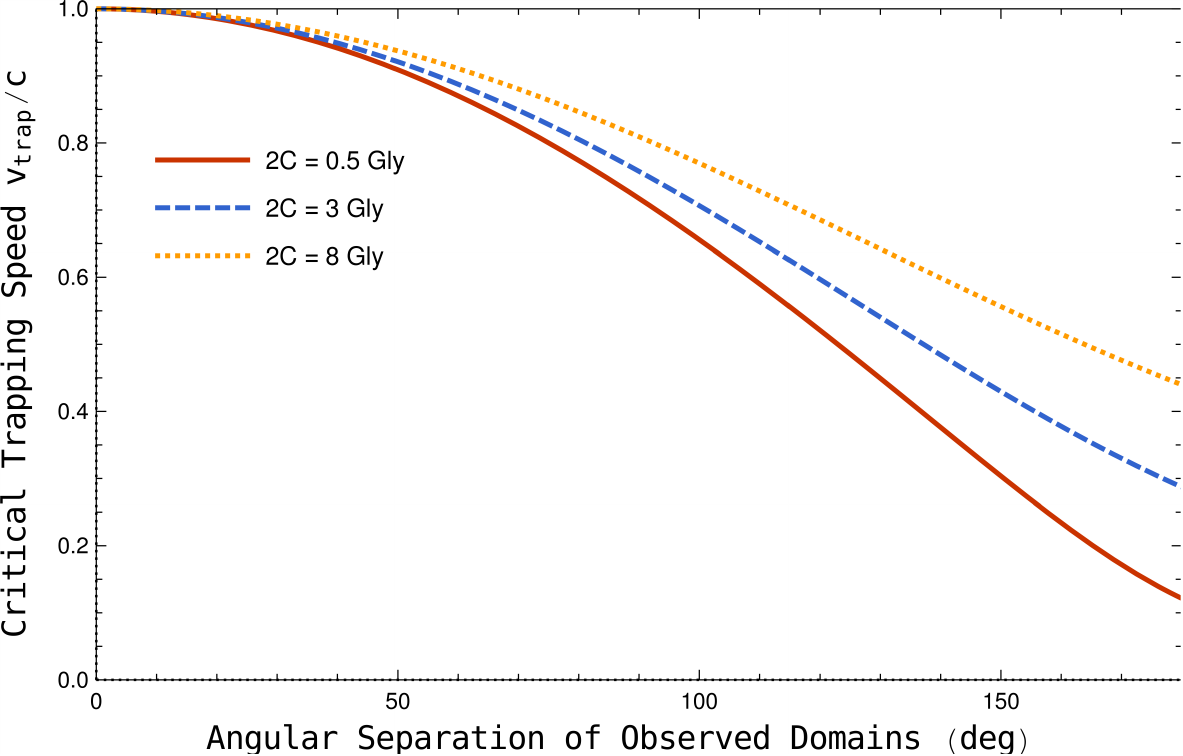}
	\caption{Dependence of the critical trapping speed on the angular separation of two visible expanding civilizations.}
\end{figure}

We express the maximum expansion radii again as $r_{1a}(\infty) = r_{1b}(\infty) = v X_2 + 2 v C$ and $r_{2}(\infty) = v X_2$, where $X_2$ is again taken to be $X_2 = \int_{t_0}^{\infty} \frac{1}{a(t')} \, dt'$.  The equations for three spheres intersecting at a point then allows us to solve for $v_{trap}$, giving:
\begin{eqnarray}
\resizebox{.42 \textwidth}{!}{$
v_{trap} =\frac{\sqrt{\sqrt{2} X_2 \cos \left(\frac{\theta }{2}\right) \sqrt{8 C^2+8 C X_2+X_2^2 \cos (\theta )+X_2^2}+(2 C+X_2)^2+X_2^2 \cos (\theta )}}{2 (C+X_2)}.$}
\end{eqnarray}
As expected, $v_{trap}$ decreases from $1$ (the speed of light) at $\theta = 0$ to a minimum value of $\sqrt{\frac{C}{C+X_2}}$ at $\theta = 180^{\circ}$ -- the curve is illustrated in Figure 6.  As before, if civ 2 corresponds to humanity or some other species on the cusp of cosmic expansion at $t_0$, then $X_2 \approx 16.7$ Gly for our cosmological parameters.

\begin{figure}[h]
	\centering
	\includegraphics[width=0.9\linewidth]{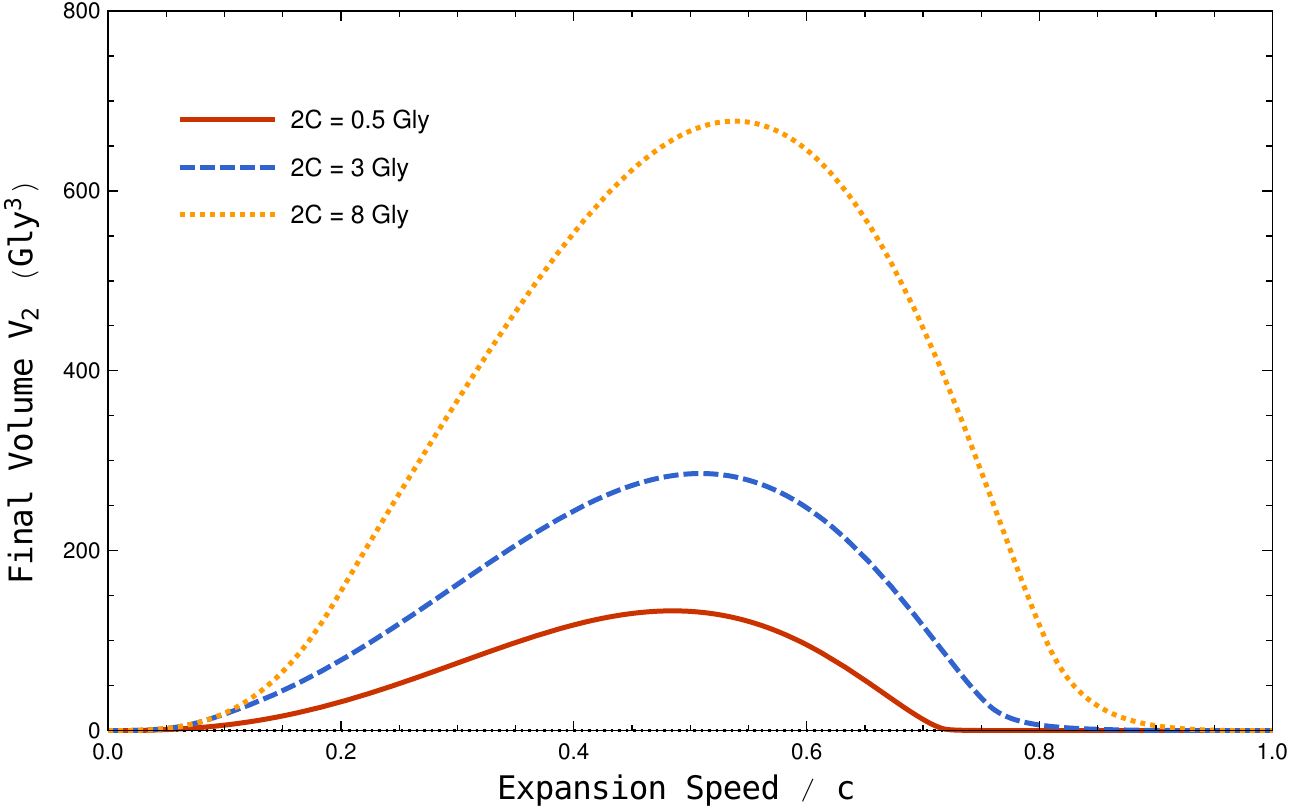}
	\caption{Final co-moving volume of civ 2 in the case of two visible domains, with an angular separation of $\theta = 90^{\circ}$.  Both observed domains appear at the same distance $2C$ from civ 2.}
\end{figure}

Though we do not have a simple volume equation, we can numerically calculate the final volume of civ 2, which appears in Figure 7, describing scenarios with a constant angular separation of $\theta = 90^{\circ}$.  The final volume is much more limited than the single-observation scenario, the optimal value for $v$ is lower, and one can see that the final volume becomes rapidly cut off as $v$ exceeds the critical trapping speed.

\section{Discussion and conclusions}

It may seem incredible that the subject of this paper can be modeled at all, when human civilizations on the Earth appear unpredictable on a much shorter timescale.  This is offset by some important features of this case, namely the large scale homogeneity of the universe, and the simplicity of extreme objects, which tend to push limits and exploit the available symmetries.  Modeling the engineering of a single solar system or tracking the migration of early hominids on the Earth is likely to be far more complex and uncertain than predicting the large-scale expansion of a highly-driven superintelligence within a homogeneous distribution of resources.  Our conclusions are the result of this kind of extreme-object simplification when symmetries are present.

If a rapidly expanding civilization is observed in a cosmological galaxy survey, humanity's potential for future cosmic expansion will be affected for nearly all plausible values of the separation distance and expansion speed.  If practical limits to expansion speed are above $\approx .75 c$, an observation means that higher limits to technology impose ever more severe limits to our future ambitions, as first-mover advantage becomes overwhelming.  Furthermore, a great separation distance does not shield us from these implications -- a factor of $\approx 2$ in the number of galaxies available to us separates a detection in nearby galaxies from a detection at the greatest plausible distance.  

These implications are amplified if more than one expanding domain is observed.  At high $v$ (expansion again above $\approx .75 c$), two detections are likely to imply that one's future will be ``trapped" between the domains of the two observed civilizations.  In high-speed scenarios, future cosmic ambitions are cut off even more rapidly than in the single-observation case.

We should note that these results have been ``best case scenarios" in the sense that we assumed all observed domains were detected at the earliest stage of expansion, although the results can be generalized with the equations of section II.  Observing a civilization in a more advanced stage of expansion is of course more limiting to the observer.  Our results are also ``best case" in the sense that we did not consider the possibility of encountering additional civilizations that were not yet visible at $t_0$.

It is also important to note that all scenarios are not equally probable.  Civilizations with an increasingly high expansion speed are increasingly unlikely to be observed, as the window to observe them becomes narrow -- one must be in the right place at the right time to see them (inside their future light cone, but just barely).  The fact that humanity is already making realistic plans to launch interstellar space probes at $.2 c$ in a time frame of $\approx 20$ years~\cite{merali2016} suggests that practical limits to expansion speed may indeed be high for advanced civilizations or superintelligence, even before we begin to consider the energy resources available to a type ii civilization for this purpose~\cite{armstrong2013}.  If we live in a universe where extremely rapid expansion is easy, present-day surveys are unlikely to detect anything, even if a substantial fraction of the universe has already been engineered by advanced life~\cite{olson2015a,olson2016a}.

In our analysis, we assumed that civilizations do not wish to share or fight one another for resources.  This is a default assumption, but it naturally raises the question of interactions between cosmological supercivilizations.   Such questions might appear hopelessly complex, but again there are a number of simplifying assumptions one can make -- for example, a superintelligence that forms a singleton with a simple utility function might behave in a very uniform and predictable manner, and be quite amenable to game theory considerations.  Our results here may regarded as a framework or a starting point for developing such ideas.

\appendix*
\section{Plausibility of the thin-boundary, constant-$v$ expansion model}
The expansion model used here, with its constant-$v$ and thin-boundary assumptions, is natural to the context of a completely homogeneous cosmology, but it is useful to see how it can emerge in the context of cosmic structure with discrete galaxies, with plausible technology and expansion-strategy assumptions.  

We consider here a case in which a home galaxy sends self-replicating spacecraft to all galaxies within some co-moving radius, $\mathcal R$.  The spacecraft are assumed to be given an initial boost with some velocity $v$, and coast until they approach their destination, i.e. they follow a geodesic.  Having arrived at their destination and finding solar systems with suitable resources, the spacecraft begin reproducing, and sending out the next generation of spacecraft to every galaxy within the same co-moving radius $\mathcal R$.

In this form of expansion, the ``effective" distance from the origin to any given galaxy is the shortest galaxy-to-galaxy path distance from the origin, under the constraint that no single jump exceeds $\mathcal R$.  Since typical intergalactic distances are Mly, the time required for all mission stages like boost, deceleration, and reproduction will account for a tiny fraction of the path travel time (assuming any known or proposed technology), since they will be dwarfed by orders of magnitude by the time required to coast between galaxies.

The assumption of a geodesic flight between galaxies is also well-approximated by a constant velocity in the co-moving frame, for reasonable parameters.  As an example, assuming an initial boost of $.1 c$ and a jump distance of $\mathcal R = .1$ Gly, the fractional difference in travel time between the constant-$v$ trajectory and the geodesic path will amount to a fraction of a percent (for a present-day launch), and the approximation gets better with higher boost speeds and shorter $\mathcal R$.  Since the velocity is re-set at each galaxy, the fractional error will not compound for long, multi-jump voyages.

The above means that the main consideration in describing the expansion of the frontier will be the distribution of shortest-path distances from the home galaxy, which is determined by $\mathcal R$ and the distribution of galaxies in space.  To obtain a quantitative description, a number of approaches might be taken.  One could use actual galaxy position data, e.g. from the Sloan Digital Sky Survey, or one could use simple models of the statistical galaxy distribution, or one could use existing data from large-scale simulations.  Here, we opt for the latter approach, using $z=0$ galaxy position data from the Millenium Run~\cite{springel2005}, which has been processed for detailed modeling of galaxy clustering and cosmic structure~\cite{croton2006}.  This approach was chosen because it sidesteps a number of systematic errors that are likely to arise in converting survey redshift data (for small values of $z$) directly into a galaxy position map.

\begin{figure}
	\centering
	\includegraphics[width=0.9\linewidth]{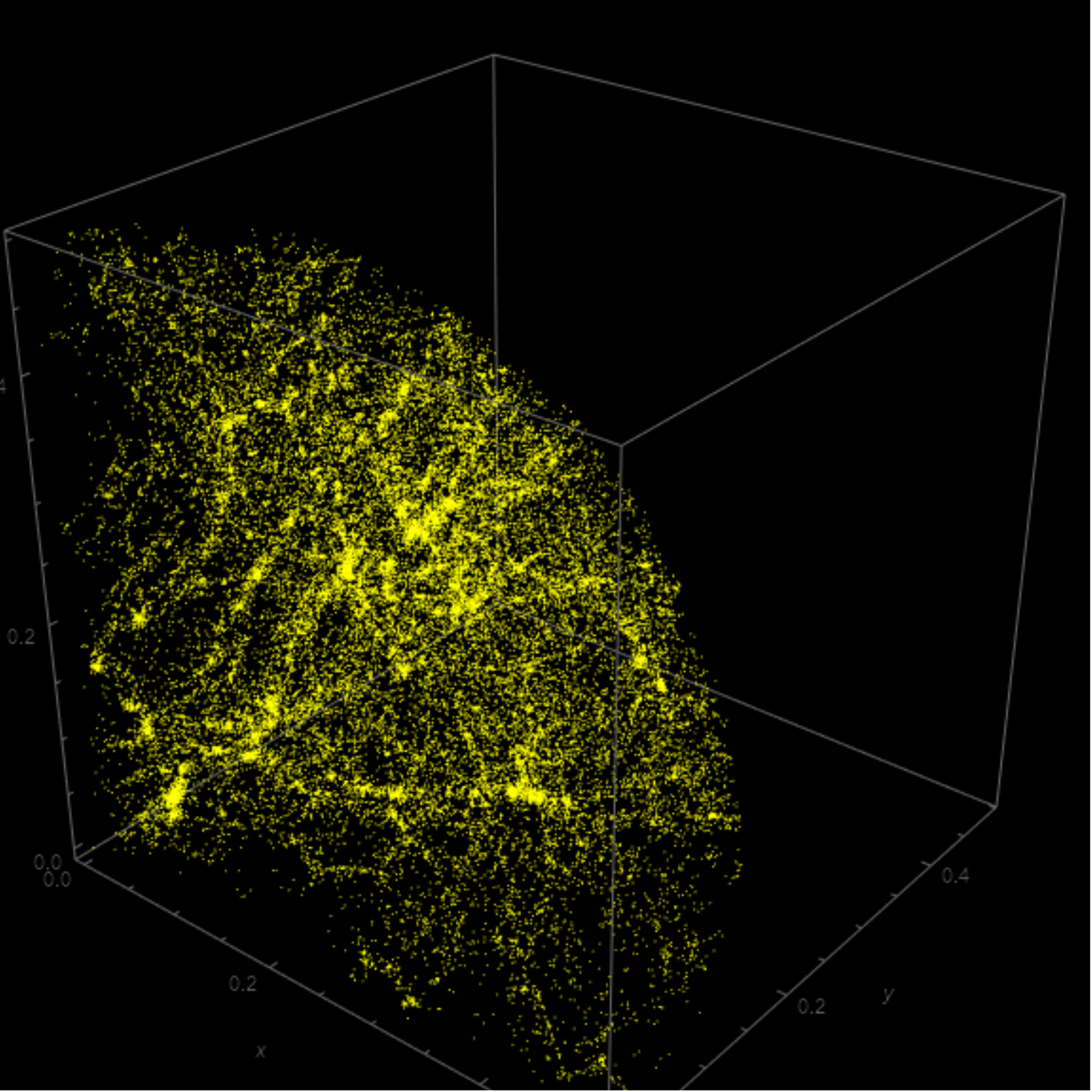}
	\caption{Set of 44,215 galaxy coordinates from the Millennium Simulation project, within $.5$ Gly of a ``home galaxy" located at the coordinate origin (bottom-left corner).}
\end{figure}

The galaxy position data available from~\cite{croton2006} is obtained in co-moving, Cartesian coordinates, in a coordinate box with each side larger than 2 Gly. The coordinate origin is one corner of the box, though the simulation used periodic boundary conditions to eliminate edge effects.  Here, we place an additional ``home" galaxy at the coordinate origin, and consider the path distance to galaxies within $.5$ Gly co-moving distance, where the longest allowed jump distance is given by $\mathcal R = 60$ Mly.  Using the coordinate corner means that expansion takes place through one eighth of a sphere -- this is one way to keep numerical calculations practical for a personal computer.  This set corresponds to 44,215 galaxies.

\begin{figure}[H]
	\centering
	\includegraphics[width=0.9\linewidth]{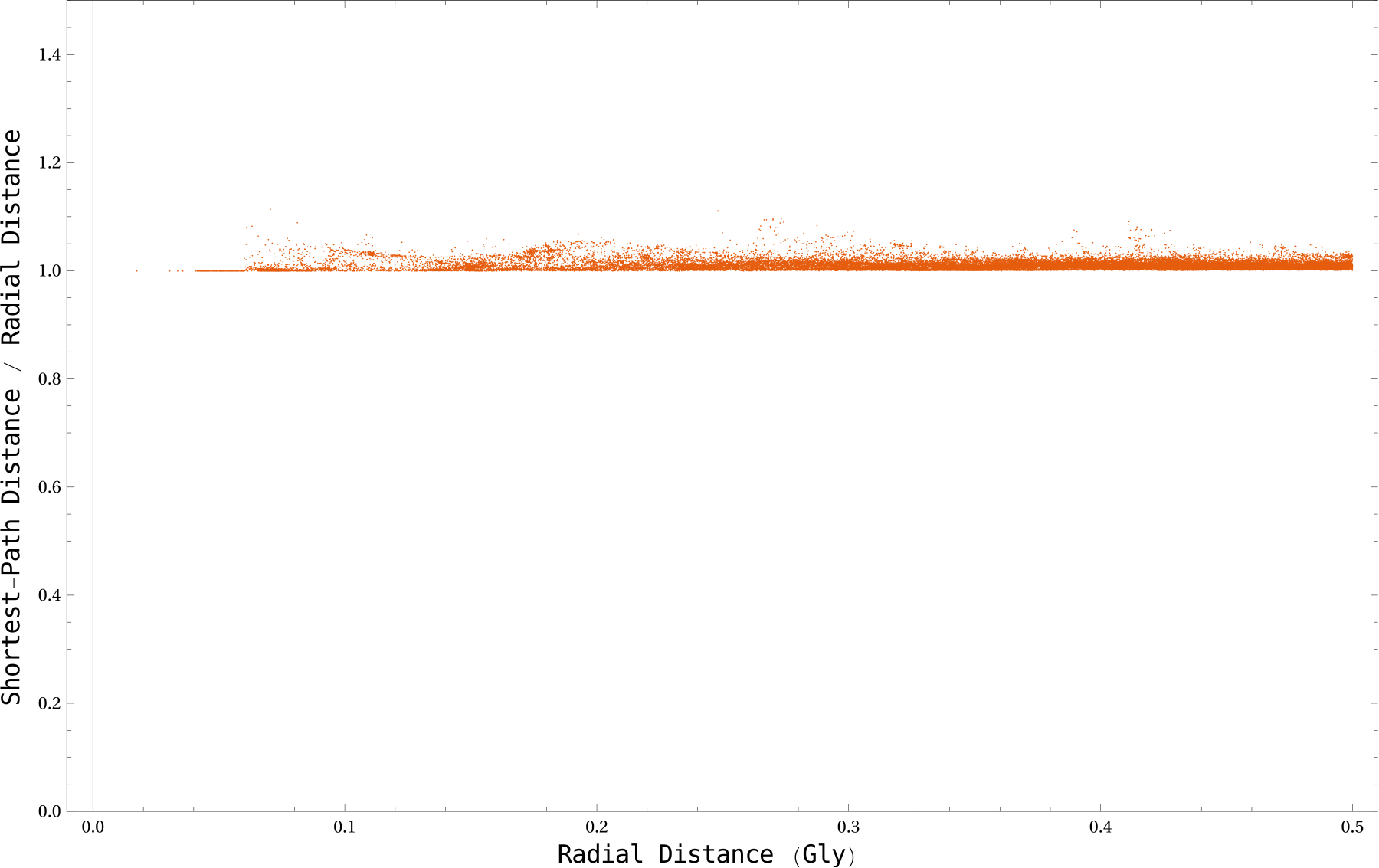}
	\caption{Ratio of path-distance to direct radial distance from the home galaxy to 44,215 galaxies, as a function of radial distance.}
\end{figure}

The galaxy position set can be seen in Figure 8, where the cosmic filamentary/supercluster/void structure is clearly visible.  The $.5$ Gly radius of the sample was chosen to safely exceed the homogeneity scale of the universe, and the $\mathcal R = 60$ Mly cutoff for individual jumps was chosen to be comparable with cluster-to-cluster distances, while remaining short of a full supercluster or void distance scale, i.e. we wish our calculation to remain sensitive to the structure of the universe below the homogeneity scale.  There are 148 galaxies within $\mathcal R$ of the home galaxy in this set, which are reached directly in the first jump.

Using these parameters, we numerically calculate the shortest-path distance to every galaxy in the set, and illustrate the ratio of shortest-path distance to radial coordinate distance in Figure 9.  Because of the large number of data points in the figure, some spread is visible.  However, the mean ratio of the shortest-path distance to the radial coordinate distance in this sample is 1.011, with a standard deviation of 0.0085, illustrating the utility of a thin-shell, constant-$v$ expansion model as an approximation.

\begin{acknowledgements}
I am especially grateful to Toby Ord and Anders Sandberg for comments and discussions regarding the branching model of expansion described in the appendix, and its relation to percolation theory.
\end{acknowledgements}

\bibliography{ref5}{}

\begin{thebibliography}{10}

\bibitem{annis1999b}
James Annis.
\newblock Placing a limit on star-fed kardashev type iii civilisations.
\newblock {\em Journal of the British Interplanetary Society}, 52:33--36, 1999.

\bibitem{armstrong2013}
Stuart Armstrong and Anders Sandberg.
\newblock Eternity in six hours: Intergalactic spreading of intelligent life
  and sharpening the fermi paradox.
\newblock {\em Acta Astronautica}, 89:1--13, 2013.

\bibitem{ball1973}
John~A Ball.
\newblock The zoo hypothesis.
\newblock {\em Icarus}, 19(3):347--349, 1973.

\bibitem{bostrom2012}
Nick Bostrom.
\newblock The superintelligent will: Motivation and instrumental rationality in
  advanced artificial agents.
\newblock {\em Minds and Machines}, 22(2):71--85, 2012.

\bibitem{bostrom2014}
Nick Bostrom.
\newblock {\em Superintelligence: Paths, dangers, strategies}.
\newblock Oxford University Press, Oxford, 2014.

\bibitem{croton2006}
Darren~J Croton, Volker Springel, Simon~DM White, Gabriella De~Lucia, Carlos~S
  Frenk, Liang Gao, Adrian Jenkins, Guinevere Kauffmann, JF~Navarro, and Naoki
  Yoshida.
\newblock The many lives of active galactic nuclei: cooling flows, black holes
  and the luminosities and colours of galaxies.
\newblock {\em Monthly Notices of the Royal Astronomical Society},
  365(1):11--28, 2006.

\bibitem{fogg1988}
Martyn Fogg.
\newblock Feasibility of intergalactic colonisation and its relevance to seti.
\newblock {\em Journal of the British Interplanetary Society}, 41:491--496,
  1988.

\bibitem{gertz2016}
John Gertz.
\newblock Reviewing meti: A critical analysis of the arguments.
\newblock {\em arXiv preprint arXiv:1605.05663}, 2016.

\bibitem{griffith2015}
Roger Griffith, Jason Wright, Jessica Maldonado, Matthew~S Povich, Steinn
  Sigurdjsson, and Brendan Mullan.
\newblock The {\^g} infrared search for extraterrestrial civilizations with
  large energy supplies. iii. the reddest extended sources in wise.
\newblock {\em The Astrophysical Journal Supplement Series}, 217(2):25, 2015.

\bibitem{hart1975}
Michael~H Hart.
\newblock Explanation for the absence of extraterrestrials on earth.
\newblock {\em Quarterly Journal of the Royal Astronomical Society}, 16:128,
  1975.

\bibitem{jones1976}
Eric~M Jones.
\newblock Colonization of the galaxy.
\newblock {\em Icarus}, 28(3):421--422, 1976.

\bibitem{kardashev1964}
Nikolai~S Kardashev.
\newblock Transmission of information by extraterrestrial civilizations.
\newblock {\em Soviet Astronomy}, 8:217, 1964.

\bibitem{lacki2016}
Brian~C Lacki.
\newblock Type iii societies (apparently) do not exist.
\newblock {\em arXiv preprint arXiv:1604.07844}, 2016.

\bibitem{merali2016}
Zeeya Merali.
\newblock Shooting for a star.
\newblock {\em Science}, 352(6289):1040--1041, 2016.

\bibitem{olson2014}
S~Jay Olson.
\newblock Homogeneous cosmology with aggressively expanding civilizations.
\newblock {\em Classical and Quantum Gravity}, 32(21):215025, 2015.

\bibitem{olson2015a}
S~Jay Olson.
\newblock Estimates for the number of visible galaxy-spanning civilizations and
  the cosmological expansion of life.
\newblock {\em International Journal of Astrobiology}, FirstView:1--9, 4 2016.

\bibitem{olson2016a}
S~Jay Olson.
\newblock On the visible size and geometry of aggressively expanding
  civilizations at cosmological distances.
\newblock {\em Journal of Cosmology and Astroparticle Physics}, 2016(04):021,
  2016.

\bibitem{omohundro2008}
Stephen~M Omohundro.
\newblock The basic ai drives.
\newblock In {\em AGI}, volume 171, pages 483--492, 2008.

\bibitem{sagan1983}
Carl Sagan and William~I Newman.
\newblock The solipsist approach to extraterrestrial intelligence.
\newblock {\em Quarterly Journal of the Royal Astronomical Society}, 24:113,
  1983.

\bibitem{springel2005}
Volker Springel, Simon~DM White, Adrian Jenkins, Carlos~S Frenk, Naoki Yoshida,
  Liang Gao, Julio Navarro, Robert Thacker, Darren Croton, John Helly, et~al.
\newblock Simulations of the formation, evolution and clustering of galaxies
  and quasars.
\newblock {\em Nature}, 435(7042):629--636, 2005.

\bibitem{tipler1980}
Frank~J Tipler.
\newblock Extraterrestrial intelligent beings do not exist.
\newblock {\em Quarterly Journal of the Royal Astronomical Society},
  21:267--281, 1980.

\bibitem{valdes1980}
Francisco Valdes and RA~Freitas.
\newblock Comparison of reproducing and nonreproducing starprobe strategies for
  galactic exploration.
\newblock {\em British Interplanetary Society, Journal(Interstellar Studies)},
  33:402--406, 1980.

\bibitem{villarroel2016}
Beatriz Villarroel, I{\~n}igo Imaz, and Josefine Bergstedt.
\newblock Our sky now and then $-$ searches for lost stars and impossible
  effects as probes of advanced extra-terrestrial civilisations.
\newblock {\em In press, The Astronomical Journal}, 2016.

\bibitem{wright2014b}
JT~Wright, RL~Griffith, S~Sigurdsson, MS~Povich, and B~Mullan.
\newblock The g infrared search for extraterrestrial civilizations with large
  energy supplies. ii. framework, strategy, and first result.
\newblock {\em The Astrophysical Journal}, 792(1):27, 2014.

\bibitem{zackrisson2015}
Erik Zackrisson, Per Calissendorff, Saghar Asadi, and Anders Nyholm.
\newblock Extragalactic seti: The tully--fisher relation as a probe of dysonian
  astroengineering in disk galaxies.
\newblock {\em The Astrophysical Journal}, 810(1):23, 2015.

\end{thebibliography}
\bibliographystyle{plain}

\end{document}